\documentclass[12pt]{iopart}
\usepackage{iopams}  
\usepackage{timet}
\usepackage{graphicx}
\usepackage{subfigure}
\begin{document}

\title[Muon tomography and gravimetry joint inversion]{Joint inversion of muon tomography and gravimetry - a resolving kernel approach}

\author[K. Jourde  et al.]
{Kevin Jourde$^1$, Dominique Gibert$^{1,2}$, Jacques Marteau$^3$}

\address{$^1$ Institut de Physique du Globe de Paris, Sorbonne Paris Cit\'e, Univ Paris Diderot, UMR 7154 CNRS, Paris, France \\
$^2$ G\'eosciences Rennes, Univ Rennes 1, UMR 6118 CNRS, Rennes, France \\
$^3$ Institut de Physique Nucl\'eaire de Lyon, Univ Claude Bernard, UMR 5822 CNRS, Lyon, France}
\ead{jourde@ipgp.fr, dominique.gibert@univ-rennes1.fr, marteau@ipnl.in2p3.fr}
\vspace{10pt}
\begin{indented}
\item[]September 2014
\end{indented}

\begin{abstract}
Both muon tomography and gravimetry are geophysical methods that provide information on the density structure of the Earth's subsurface. Muon tomography measures the natural flux of cosmic muons and its attenuation produced by the screening effect of the rock mass to image. Gravimetry generally consists in measurements of the vertical component of the local gravity field. Both methods are linearly linked to density, but their spatial sensitivity is very different. Muon tomography essentially works like medical X-ray scan and integrates density information along elongated narrow conical volumes while gravimetry measurements are linked to density by a 3-dimensional integral encompassing the whole studied domain. We develop the mathematical expressions of these integration formulas -- called acquisition kernels -- to express resolving kernels that act as spatial filters relating the true unknown density structure to the density distribution actually recoverable from the available data. The resolving kernels provide a tool to quantitatively describe the resolution of the density models and to evaluate the resolution improvement expected by adding new data in the inversion. The resolving kernels derived in the joined muon/gravimetry case indicate that gravity data are almost useless to constrain the density structure in regions sampled by more than two muon tomography acquisitions. Interestingly the resolution in deeper regions not sampled by muon tomography is significantly improved by joining the two techniques. Examples taken from field experiments performed on La Soufri\`ere of Guadeloupe volcano are discussed.
\end{abstract}

%
%
%
%
%

\section{Introduction}

\begin{figure}
\centering
\subfigure[tomography coverage] {
	\includegraphics[width=0.48\textwidth]{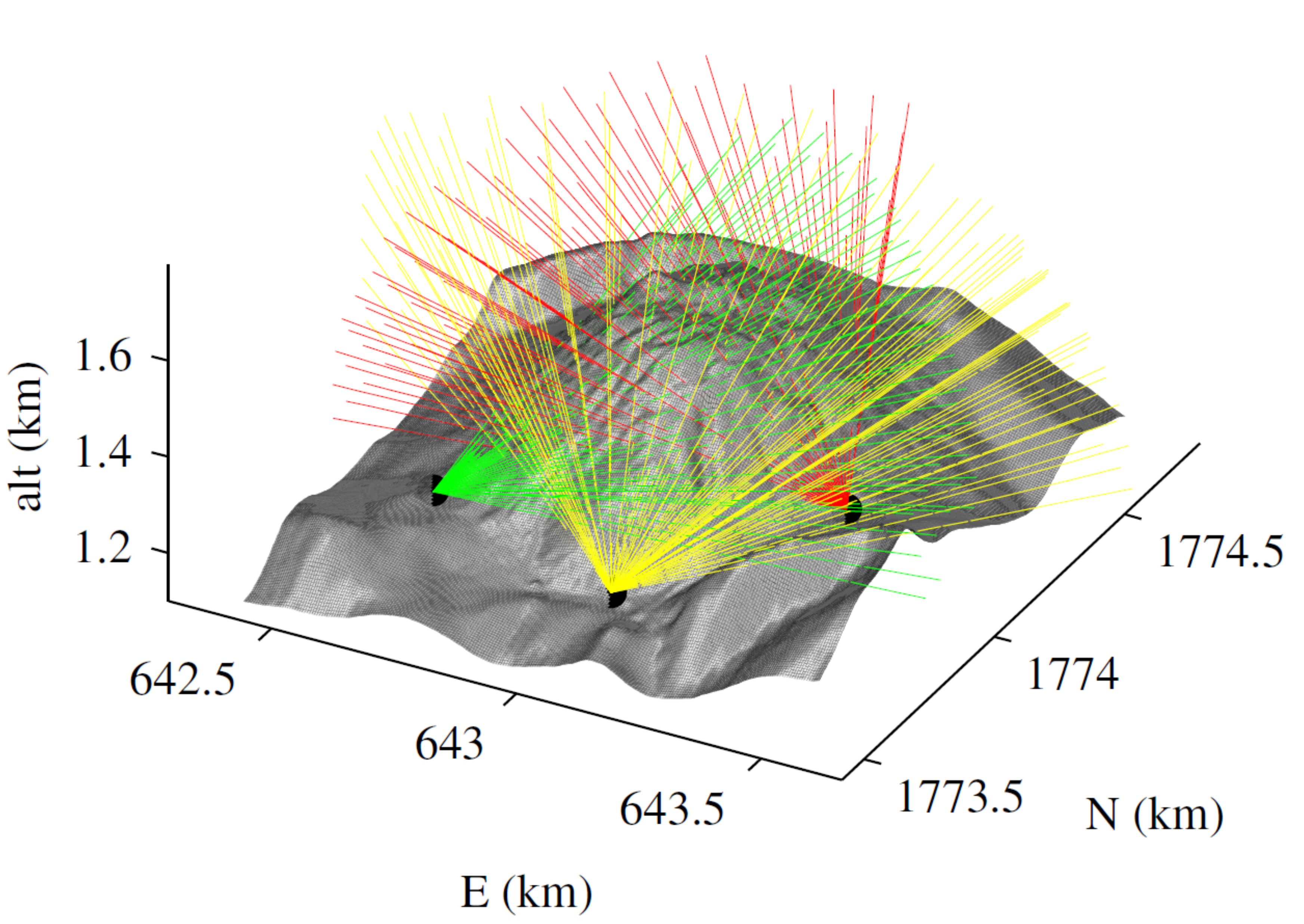}
	\label{soufriereCoverage_a}
}
\subfigure[gravimetry coverage] {
	\includegraphics[width=0.48\textwidth]{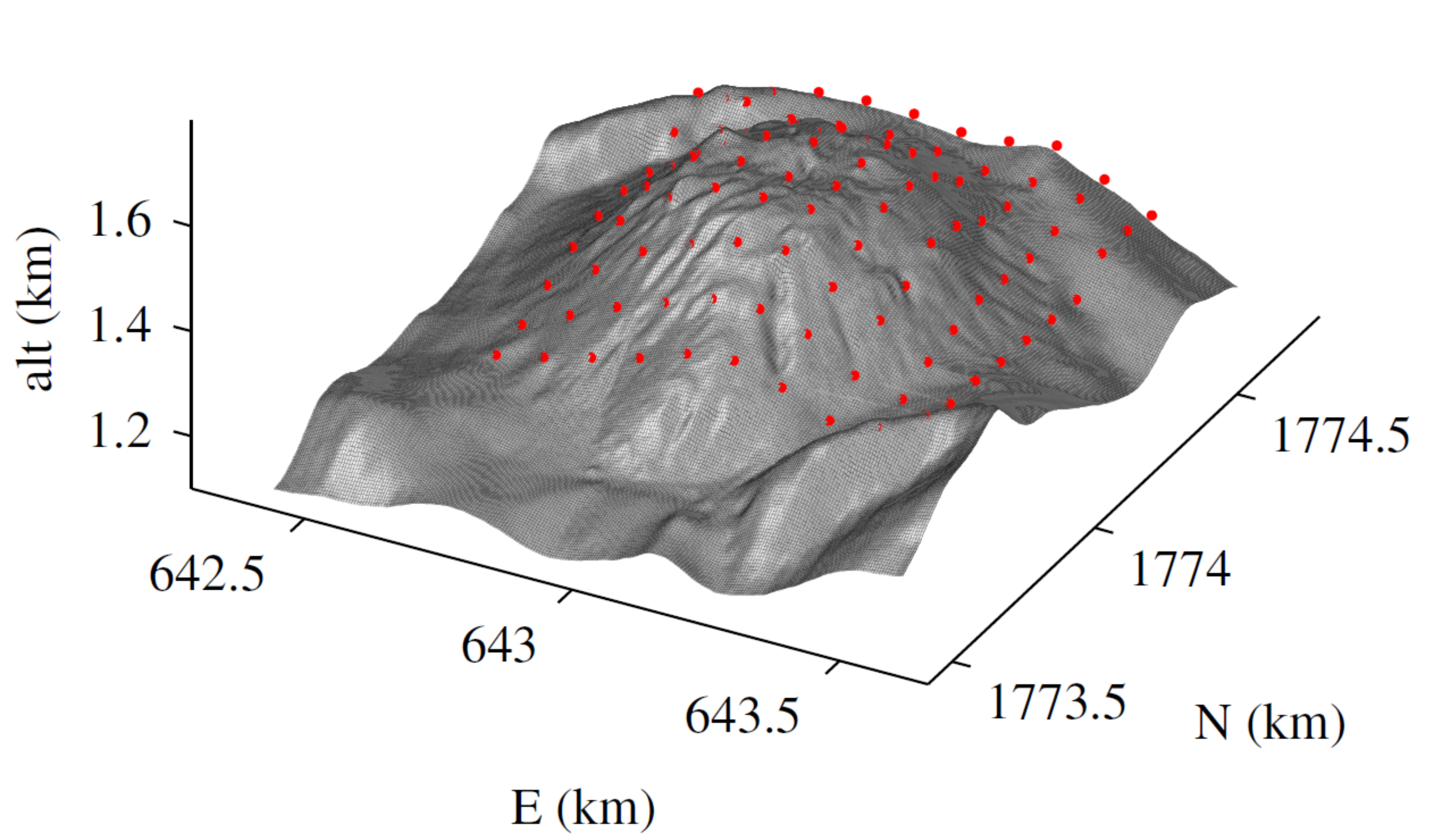}
	\label{soufriereCoverage_b}
}
\caption{\textit{(a):} muon tomography data coverage on the Soufri\`ere of Guadeloupe. The lines represent the observation axes of the muon telescope when located at Rocher Fendu (red), Ravine Sud (yellow) and Savane \`a Mulets (green). \textit{(b):}  the red dots represent the positions at which we simulated gravimetry measurements}
\label{soufriereCoverage}
\end{figure}

Muon tomography is a geophysical method that offers a new way to determine the density of large rock volumes by measuring their screening effect on the natural cosmic muons flux crossing the rock volume to probe (e.g. Nagamine 2003, see Tanaka 2013 for a brief review). The small cross-section of muons in ordinary matter (Barrett et al. 1952) allows the hard component of the muon spectrum (Tang et al., 2006; Gaisser \& Stanev 2008) to cross hectometers, and even kilometers, of rock. Most muons crossing the rock volume have a negligible scattering relative to the instrument angular resolution and travel along straight trajectories ranking muon tomography among the class of straight-ray scanning imaging methods (Marteau et al., 2011). In practice, muon tomography is performed by using a series of pixelated particle detectors that allow to determine the trajectories of the muons passing through the rock body. Portable field telescopes presently used sample hundredths of directions and allow to scan an entire volcano from a single view-point in a couple of weeks (Fig.~\ref{soufriereCoverage}). By counting the number of muons passing through the target, the attenuation onto the incident muon flux is determined for each sampled direction and used to produce a radiography of the object opacity (expressed in $\mathrm{g}.\mathrm{cm}^{-2}$) or of average density along ray-paths if the object geometry is known.

Since the pioneering works by Nagamine et al. (1995a,b) and Tanaka et al. (2001), recent studies illustrate the interest of the method to image spatial and temporal variations of the density inside volcanoes (Tanaka et al. 2005, 2007a,b, 2008, 2009a,b, 2013; Gibert et al., 2010; C\^arloganu et al., 2012; Lesparre et al. 2012c; Shinohara \& Tanaka, 2012; Portal et al., 2013). When compared with the relatively large number of publications devoted to qualitative applications of muon tomography, only a small number of studies address methodological issues and quantitative assessments of the method. Lesparre et al. (2010) establish a feasibility formula where the achievable density resolution is related to the measurement duration (i.e. time resolution), the total apparent rock thickness (i.e. total opacity) and the telescope acceptance (i.e. the detection capacity of the matrices). The feasibility formula writes as an inequality and gives practical hints to design field experiments and evaluate which density heterogeneities can be resolved inside a given geological target, for a given amount of time and a given telescope. In a more recent study, Jourde et al. (2013) present experimental evidences of a flux of upward-going particles that occurs in certain field conditions. These particles have trajectories parallel to those of the muons emerging from the rock body to radiography but  they travel through the telescope from rear to front. These upward-going particles may constitute a huge Poissonian noise that could strongly alter the radiographies. Jourde et al. (2013) give practical recommendations for choosing experimental sites likely to give an as best as possible signal-to-noise ratio, and they also puts strong constrains on the time resolution of the electronic detection chain necessary to statistically recognize particles coming from the rear face of the telescopes.

In the present study, we extend our methodological work by quantitatively examining how the combination of muon tomography and gravity data may improve the reconstruction of the density distribution inside geological bodies. Studies combining muon data and gravity measurements remain scarce, and we emphasize the early study by Caffau et al. (1997) who compared muon tomography with gravity measurements. More recently, Davis \& Oldenburg (2012) and Nishiyama et al. (2014) presented joined inversions of gravity data and muon tomography using a straightforward linear regularized inversion based on block models. Combining density radiographies and gravity measurements is a quite natural intention since both methods provide information directly related to the density distribution. However, both methods sample the density structure in very different ways, and the scanner-like principle of muon tomography confers to this method a high-resolution that could question the interest to perform a costly complementary gravimetry survey. Conversely, muon tomography has limitations due to difficulties to install muon telescopes at a large number of places around the geological body to image. Consequently the number of opacity radiographies available to perform the tomography reconstruction is always small with, eventually, wide uncovered angular sectors that make the inverse Radon transform ill-posed to perform the tomography reconstruction. Another strong limitation of muon tomography comes from the fact that it at best only samples the density structures located above the horizontal plane passing through the telescope location.

Considering these issues, gravity data may bring useful information to better constrain the recovery of the density distribution, and it is the purpose of the present study to quantitatively document the benefits of a joint inversion of gravity and muon data to recover the density structure. In order to obtain results independent of any particular parametrization (e.g. block discretization), we adopt a formulation in terms of resolving kernels whose expression only depend on the geometrical properties of the data acquisition (i.e. locations of measurement points and telescope acceptance functions). We begin by establishing the relationships between the density structure and both muon tomography and gravity data. Next, we derive the resolving kernels respectively corresponding to individual gravity and muon inversions and to joint gravity--muon inversion. The resolving kernels translate the information contained in the data into information concerning the density structure. We conclude the article with examples taken from real field experiments conducted on La Soufri\`ere of Guadeloupe to illustrate the practical interest to combine muon radiographies and gravity measurements. The muon tomographies experiments were already described in various articles (Lespare et al., 2012 and Jourde et al. 2013). Three sites called Ravine Sud, Rocher Fendu and Savane \`a Mulets were explored and are represented on Fig.~\ref{soufriereCoverage_a}. The gravimetry survey is currently running, for the purpose of this article we simulated one hundred measurements regularly spaced on a grid that covers the dome (Fig.~\ref{soufriereCoverage_b}).


\section{The sampling of the density distribution by muon tomography and gravimetry}

Here, we recall the main formula relating the density distribution to the data, i.e. fluxes of muons and gravity measurements. In the inverse problem framework, these formula describe the forward problem for each method. In the remaining, we suppose that the muon data have been cleaned from perturbing effects such as upward going fluxes of particles as described in Jourde et al. (2013).

\subsection{Muon tomography}
\label{section_muon_tomography}

\begin{figure}[h]
\begin{center}
\includegraphics[width=0.7\linewidth]{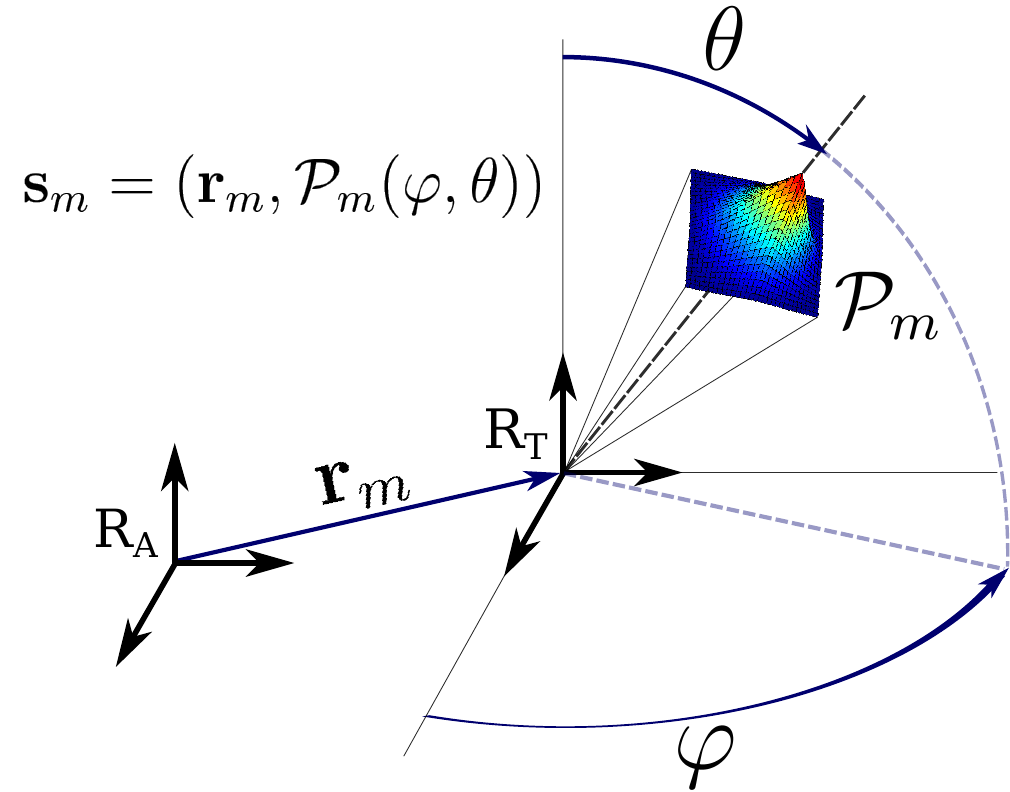}
\end{center}
\caption{Muon tomography reference frame and notations. $R_A$ and $R_T$ respectively are the absolute orthonormal and the instrument reference frames. An observation axis $\mathbf{s}_m=\left(\mathbf{r}_m,\mathcal{P}_m(\varphi,\theta)\right)$ is represented with $\mathbf{r}_m$ the vector that localizes the telescope position and $\mathcal{P}_m$ its acceptance pattern (restrained to the solid angle $\Omega_m$). The spherical coordinates $(\varphi,\theta) $ are here localizing the steep acceptance peak mentioned in section \ref{section_muon_tomography}.}
\label{Figure_Frame}
\end{figure}

The primary information used in muon tomography consists in cosmic muons flux attenuation measurements resulting from the screening produced by the geological volume to scan. Attenuation is measured by counting the number of muons emerging from the volume for each observation axis: $\mathbf{s}_m = \left(\mathbf{r}_m,\mathcal{P}_m(\varphi,\theta) \right)$, of the telescope (Fig.~\ref{soufriereCoverage_a}). $\mathbf{r}_m$  represents the position of the telescope, $\mathcal{P}_m$ the observation axis acceptance pattern which depends on $(\varphi,\theta)$ the azimuth and zenith angles referenced at $\mathbf{r}_m$ (see Fig.~\ref{Figure_Frame}). Note that $\mathbf{r}_m$ is the same for all the observation axes on a given site. $\mathcal{P}_m$ depends on the telescope geometry and angular orientation on the site. Our standard field telescopes count $31 \times 31$ observation axes and, in a field experiment where the telescope successively occupies several places around the target, the number, $M$, of data may easily reach several hundredths. For example if we use the muon tomography data from the $3$ Soufri\`ere sites $M = 3 \times 31 \times 31$. In practice $M$ is lower as many axes point downward or above the volcano.

\begin{figure}[h] 
\begin{center}
\includegraphics[width=0.5\linewidth]{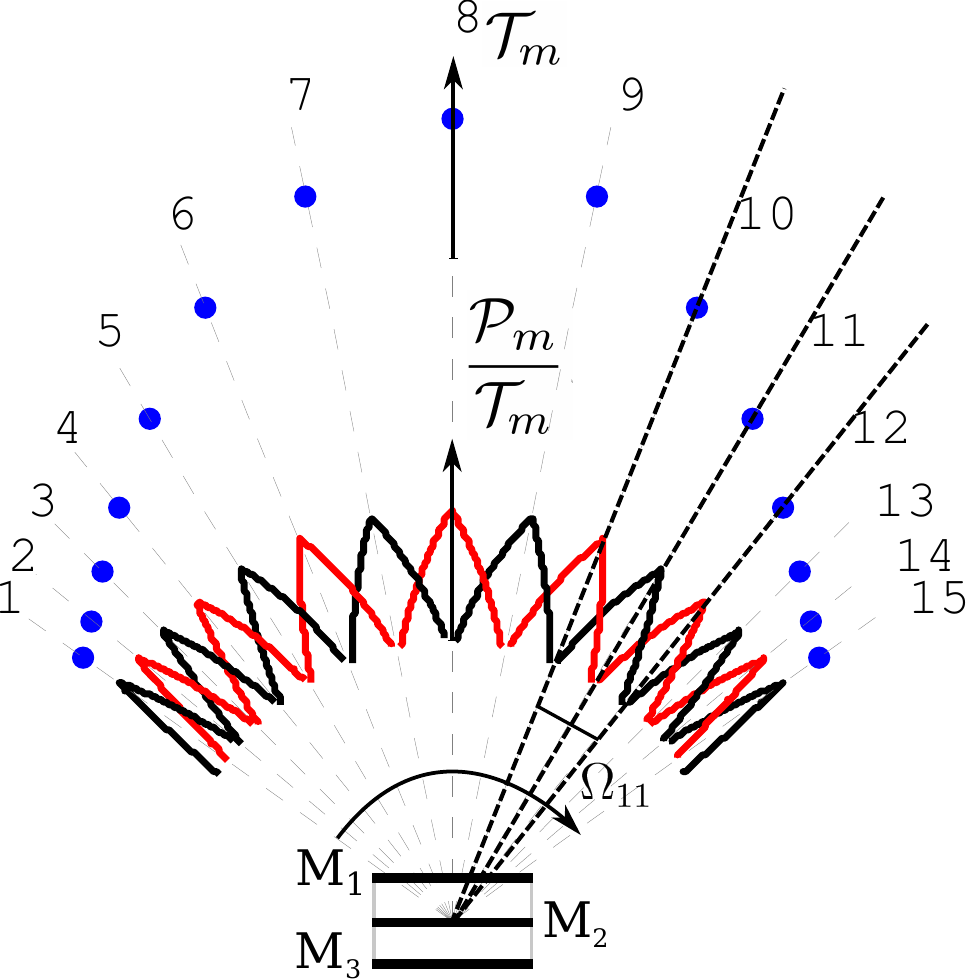}
\end{center}
\caption{Representation of the acceptance. The horizontal bars labelled $M_1$, $M_2$ and $M_3$ represent the pixelated detection matrices of the telescope. We draw $\mathcal{T}_m$ and $\mathcal{P}_m / \mathcal{T}_m$ for $15$ observation axes $\mathbf{s}_m$ symmetrically distributed on the left and right sides of the main axis of the telescope (arbitrarily indexed for $m$ going from $1$ to $15$, we do not represent all the observation axes for clarity purposes). The red and black sawtooth-like curves are a merging of the normalized intra-pixel acceptance $\mathcal{P}_m / \mathcal{T}_m$ for even and odd lines of sights respectively. The blue dots represent the $15$ discrete values of the integrated pixel acceptance $\mathcal{T}_m$ obtained by integrating each $\mathcal{P}_m$ on the unit sphere (eq.~(\ref{integratedAcceptance})). Observe that the solid angle $\Omega_m$ associated with a given $\mathbf{s}_m$ (here represented for $m=11$) overlaps the solid angle of the neighbour lines of sight. The dashed lines are plotted along the acceptance steep peaks discussed in section \ref{section_muon_tomography}.}
\label{Figure_Pixel_Acceptance}
\end{figure}

$\mathcal{P}_m$ $(\mathrm{cm}^2.\mathrm{sr}.\mathrm{rad}^{-2})$ shape depends on the detection matrices structure (see Lesparre et al. 2012a,b for further details).  It has a steep peak centered on a small solid angle region $\Omega_m$. It is identically null outside $\Omega_m$ (Fig.~\ref{Figure_Frame} and Fig.~\ref{Figure_Pixel_Acceptance}). Observe that $\mathcal{P}_m$ must not be confounded with the integrated pixel acceptance $\mathcal{T}_m$ $(\mathrm{cm}^2.\mathrm{sr})$ used for instance in Lesparre et al. (2010),

\begin{equation}
\mathcal{T}_m =  \int_{0}^{2\pi} \int_{0}^{\pi} \mathcal{P}_m \left(\varphi,\theta \right) \times \sin\left( \theta \right) \mathrm{d}\theta \mathrm{d}\varphi = \int_{\Omega_m} \mathcal{P}_m \left(\varphi,\theta \right) \mathrm{d}\Omega.
\label{integratedAcceptance}
\end{equation}

The number of muons attributed to a given line of sight actually corresponds to all muons detected in $\Omega_m$. Inside the geological volume the trajectories of these muons describe a conical volume whose apex is located at the telescope, $\mathbf{r}_m$.

The attenuation of the muon flux caused by the rock screen depends on the amount of matter encountered by the particles along their trajectories. For a given straight trajectory $\mathbf{t} = \left(\mathbf{r},\varphi,\theta \right)$ ($\mathbf{r}$ is a telescope site and $(\varphi,\theta)$ the azimuth and zenith angles referenced at $\mathbf{r}$) it is quantified by the density line integral along $\mathbf{t}$, the opacity
\begin{equation}
\label{opacity_equation}
\varrho = \int_{\mathbf{t}} \rho(\xi) ~ \mathrm{d}\xi = L \times \bar{\rho},
\end{equation}
where $\rho$ is the density, $L$ the particle path length, and $\bar{\rho}$ is the average density along $\mathbf{t}$. The differential flux associated with $\mathbf{t}$ may be expressed as a function $\delta\phi_\mathbf{t} = \frac {\partial^3 \phi} {\partial \Omega \partial S} (\varrho, \varphi, \theta) ~~ [\mathrm{s}^{-1}.\mathrm{cm}^{-2}.\mathrm{sr}^{-1}]$ that accounts for the muon flux that reaches the instrument. Then the measured flux $\phi_m$ for the $m^\mathrm{th}$ line of sight relates to the opacity \textit{via} the relationship,
\begin{equation}
\label{muonfluxgeneral}
\phi_m =  \mathcal{T}_m^{-1}\int_{\Omega_m} \mathcal{P}_m \left(\varphi,\theta \right) \times \delta\phi_{\mathbf{t}}(\varrho, \varphi, \theta) \mathrm{d}\Omega.
\end{equation}
Note that the integration is restricted to the small solid angle $\Omega_m$ because of the compact support of the observation axis acceptance $\mathcal{P}_m$ (Fig.~\ref{Figure_Pixel_Acceptance}).

$\delta\phi_{\mathbf{t}}$ is not linearly related to $\varrho$, however for small opacity fluctuations we assume that $\delta\phi_{\mathbf{t}}$ may be approximated by its first order development around the local average density, $\rho_0 (\mathbf{r})$. $\rho_0 (\mathbf{r})$ is the prior density model of the geological structure. For a given path $\mathbf{t}$ it reads,
\begin{equation}
\delta\phi_{\mathbf{t}} \left( \varrho \right) = \delta\phi_{\mathbf{t}} \left( \varrho_0 \right) + \left(\varrho - \varrho_0 \right) \times \frac{\mathrm{d} \delta\phi_{\mathbf{t}}}{\mathrm{d}\varrho}\left( \varrho_0 \right) + o \left( \varrho \right),
\end{equation}
where $\varrho_0 = \int_{\mathbf{t}} \rho_0(\xi) ~ \mathrm{d}\xi $. Rearranging the terms and letting $\alpha_{\mathbf{t}} = \frac{ \mathrm{d} \delta\phi_{\mathbf{t}}}{\mathrm{d}\varrho}\left(\varrho_0 \right)$, we obtain,
\begin{equation}
\delta\phi_{\mathbf{t}} \left( \varrho \right) - \delta\phi_{\mathbf{t}} \left( \varrho_0 \right) \approx \alpha_{\mathbf{t}} \int_{\mathbf{t}} \left[\rho(\xi) - \rho_0(\xi) \right] \mathrm{d}\xi.
\label{linear_dev}
\end{equation}
Inserting eq. \ref{linear_dev} into eq. \ref{muonfluxgeneral} we get the approximate equation
\begin{equation}
\phi_m -\phi_0 \approx
\int_{\Omega_m} \mathrm{d}\Omega \int_{\mathbf{t}} \frac{\mathcal{P}_m \left( \varphi,\theta \right)}{\mathcal{T}_m} \times \alpha_{\mathbf{t}} \times \left[\rho \left( \xi \right) - \rho_0(\xi) \right]  \mathrm{d}\xi 
\label{muonkernelSpherical}
\end{equation}
where $\phi_0 = \phi_m (\rho_0)$ is the flux corresponding to the prior density model $\rho_0(\mathbf{r})$ and $\mathbf{t} = \left(\mathbf{r}_m,\varphi,\theta \right)$.

In the remaining of the present paper, we shall use the centred and normalized flux,
\begin{equation}
\tilde{\phi}_m = \frac{\phi_m - \phi_m\left( \rho_0 \right)}{\phi_m \left( \bar{\rho}_{\min} \right) - \phi_m \left( \bar{\rho}_{\max} \right)} = \frac{\phi_m - \phi_0}{C_{\phi,m}},
\label{normalizedCentredFlux}
\end{equation}
where $\bar{\rho}_{\min}$ and $\bar{\rho}_{\max}$ are expected extreme values of the density.

\subsection{Gravimetry}

Gravimetry aims to estimate the gravity field generated by surrounding objects measuring locally the vertical acceleration they produce. The vertical acceleration $g$ is directly related to the density spatial distribution through the Newton law:
\begin{equation}
g_n = G \int_V \frac{\left(\mathbf{r}_{n} - \mathbf{r}\right) \centerdot \mathbf{e}_z}{\Vert\mathbf{r}_{n} - \mathbf{r}\Vert^3} \times \rho(\mathbf{r}) \mathrm{d}\mathbf{r}
\label{graviKernel01}
\end{equation}
where the vector $\mathbf{r}_{n}$ represents the location of the $n^\mathrm{th}$ measurement point (in our example $n$ runs from $1$ to $100$). As for muon tomography we use the normalized gravity anomaly $\tilde{g}_n$ defined as
\begin{equation}
\eqalign{
\tilde{g}_n & = \frac{g_n-g_n(\rho_0)}{C_{g,n}}  \cr
C_{g,n} & = |g_n(\bar{\rho}_{\min})-g_n(\bar{\rho}_{\max})|
}
\label{normalizedGravity}
\end{equation}

\section{Resolving kernel approach}

\subsection{The acquisition kernels}

We define $X$, the space that contains the set of continuous $L^2$ functions going from $\mathbb{R}^3$ into $\mathbb{R}$. The 3D density distribution $\rho$ belongs to $X$ and it is related to the muon flux measurements, $\tilde{\phi}_m$, and to the gravity data, $\tilde{g}_n$, through the action of acquisition kernels $\mathcal{G}$ and $\mathcal{M}$ which also belong to $X$. This reads,
\begin{eqnarray}
\tilde{\phi}_m & = { \langle \mathcal{M}_m , \rho - \rho_0 \rangle }_X,~~ m=1, \cdots, M \label{acquisitionKernel_1a} \\
\tilde{g}_n & = { \langle \mathcal{G}_n , \rho - \rho_0 \rangle }_X,~~~~~ n=1, \cdots, N, \label{acquisitionKernel_1b}
\end{eqnarray}
where ${\langle \cdot , \cdot \rangle}_X$ is $X$ inner scalar product, and $M$ and $N$ are respectively the number of muon tomography and gravimetry data. From eq. (\ref{muonkernelSpherical}) (\ref{normalizedCentredFlux}) (\ref{graviKernel01}) and (\ref{normalizedGravity}) we obtain explicit expressions for $\mathcal{M}$ and $\mathcal{G}$,
\begin{eqnarray}
\mathcal{M}_m (\mathbf{r}) & =  \frac{\mathcal{P}_m \left( \varphi,\theta \right)}{\mathcal{T}_m} \times \frac{\alpha_{\mathbf{t}}}{C_{\phi,m} ~ \xi^2}, \label{acquisitionKernel_2a} \\
\mathcal{G}_n (\mathbf{r}) & = \frac{G}{C_{g,n}} \times \frac{\left(\mathbf{r}_n - \mathbf{r}\right)}{\Vert \mathbf{r}_n - \mathbf{r}\Vert^3} \centerdot \mathbf{e}_z. \label{acquisitionKernel_2b}
\end{eqnarray}

\begin{figure}[h]
\centering
\subfigure[gravimetry acquisition kernel, $\mathcal{G}$]{
	\includegraphics[width=0.47\textwidth]{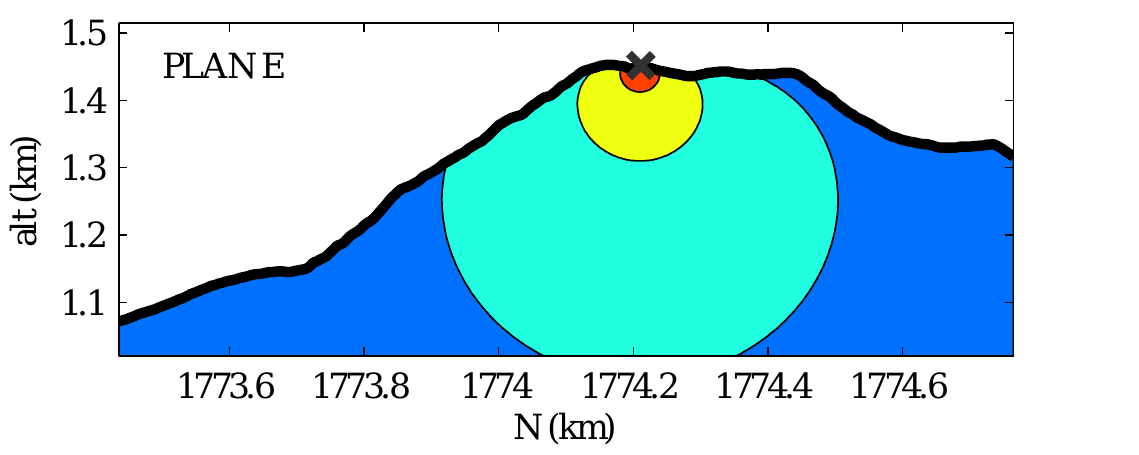}
	\label{acquisitionKernels_a}
}
\hspace{-10mm}
\subfigure[tomography acquisition kernel, $\mathcal{M}$]{
	\includegraphics[width=0.47\textwidth]{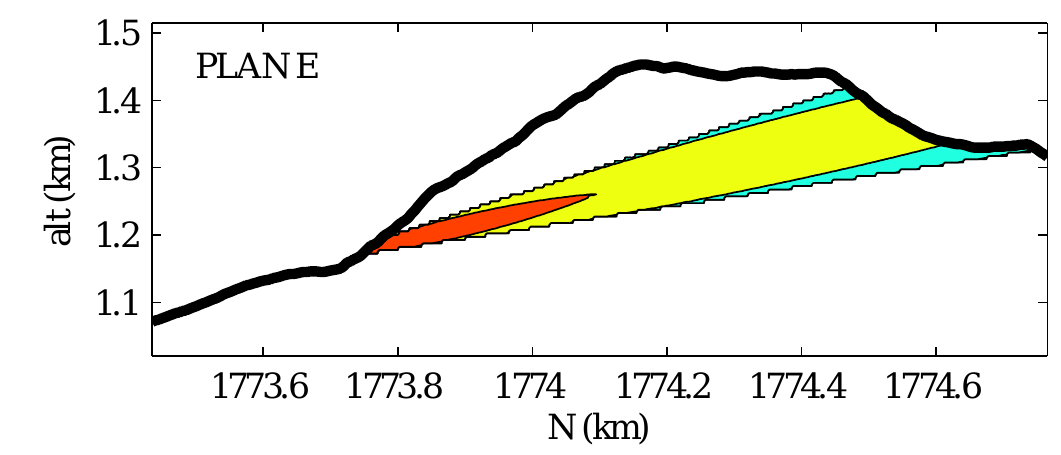}
	\label{acquisitionKernels_b}
}
\subfigure[] {
	\includegraphics[width=0.06\textwidth]{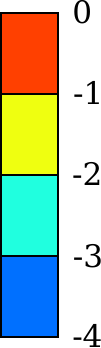}
}
\caption{Acquisition kernel of a gravimetric measurement, \textit{(a)}, and of a tomographic measurement (for one observation axis) on the right, \textit{(b)}. The acquisition kernels are normalized with reference to their maximum value and printed in a $\mathrm{log_{10}}$ scale.}
\label{acquisitionKernels}
\end{figure}

Observe that the $1/\xi^2$ term in eq.~(\ref{acquisitionKernel_2a}) comes from the spherical coordinates elementary volume expression, $\xi^2 \mathrm{d}\xi \mathrm{d}\Omega$, inserted in eq.~(\ref{muonkernelSpherical}). Examples of acquisition kernels are plotted in Fig.~\ref{acquisitionKernels} and discussed in sections \ref{section_gravimetry_kernels} and \ref{section_muon_tomography_kernels}. The $X$ structure allows to introduce prior information into the problem. For instance, the classical inner product of $L^2$ continuous functions,
\begin{equation}
\label{acquisitionKernel_3}
{\langle f,g \rangle}_X = \int_V f(\mathbf{r}) \times g(\mathbf{r}) \mathrm{d}\mathbf{\mathbf{r}},
\end{equation}
can be replaced by the weighted inner product,
\begin{equation}
\label{X_scalar_Product}
{\langle f,g \rangle}_X = \int_V \int_V w(\mathbf{r'},\mathbf{r''}) \times f(\mathbf{r'}) \times g(\mathbf{r''}) \mathrm{d}\mathbf{\mathbf{r'}} \mathrm{d}\mathbf{\mathbf{r''}},
\end{equation}
where the weight function $w$ ($w(\mathbf{r'},\mathbf{r'}) > 0$, $w(\mathbf{r'},\mathbf{r''}) = w(\mathbf{r''},\mathbf{r'})$),  plays the role of a covariance function that may be used to neglect the impact of the free air zone around the studied structure for gravimetry and muon tomography (see eq.~(\ref{resolvingKernel_5c}) and its comment below). It may also serve to introduce a correlation length for the density variations.

\subsection{The resolving kernel}

The 3D density distribution, $\hat{\rho}(\mathbf{r})$, obtained by solving the set of linear equations (\ref{acquisitionKernel_1a},\ref{acquisitionKernel_1b}) is a degraded version of the true density distribution, $\rho(\mathbf{r})$, both because of the limited number of data available and of the filtering (i.e. blurring) effect of the acquisition kernels. In the remaining, we shall use the set of undifferentiated acquisition kernels
\begin{equation}
\{ \mathcal{\zeta}_k \} = \{ \mathcal{G}_n \} \cup \{ \mathcal{M}_m \} \;\; k = 1,\cdots,K=M+N,
\label{acquisitionKernelFamily}
\end{equation}
and the set of undifferentiated normalized data,
\begin{equation}
\{ d_k \} = \{ \tilde{g}_n \} \cup \{ \tilde{\phi}_m \} \;\; k = 1,\cdots,K=M+N.
\label{normalizedDataFamily}
\end{equation}

We now formulate the inverse problem in the framework of functional spaces where the family of acquisition kernels constitutes a set of generating functions of a subspace $X_K$ of $X$ of dimension $K$ (Tarantola \& Nercessian, 1984; Bertero et al., 1985). This implicitly assumes that the $\mathcal{\zeta}_k$ are linearly independent with respect to the retained inner product, i.e. no acquisition kernel can be written as a linear combination of the other kernels. The noticeable instances where this important assumption is not satisfied correspond to situations where several data have been acquired identically, i.e. either at the same location for gravity measurements or with the same position and orientation of the telescope for muon tomography. In such cases the dimension of $X_K$ is reduced since the redundant data may be merged (i.e. averaged) into a single one.

The best density distribution that can be recovered through the inversion process (it is the best because it takes all the information contained in the data and makes the less hypotheses about $X_K$ complementary subspace) is a linear combination of the generating functions,
\begin{equation}
\label{resolvingKernel_1}
\hat{\rho}(\mathbf{r}) - \rho_0 (\mathbf{r}) = \sum_{k=1}^{K} a_k \times \zeta_k(\mathbf{r}).
\end{equation}

The components $a_k$ of eq.~(\ref{resolvingKernel_1}) are obtained by minimizing the quadratic distance $\epsilon_Y$ between the data and the corresponding values given by the density model,
\begin{equation}
\label{resolvingKernel_1b}
\epsilon_Y  = {\Vert \langle \{ \zeta_k \} , \rho-\rho_0 \rangle_X - \{ d_k \} \Vert}_Y = \sum_{k=1}^K W_k \times ( \langle \zeta_k , \rho-\rho_0 \rangle_X - d_k )^2
\end{equation}
$Y$ is the weighted Euclidean space that contains the measurements. The weights $W_k$ permits to introduce prior information about the measurements quality. It is possible to introduce crossed terms $W_{ij}$ if the measurements are not independent, but it is not the case here. We get
\begin{equation}
\label{resolvingKernel_2}
a_k = \sum_{j = 1}^K W_j \times S^{k,j} \times \langle \zeta_j , \rho-\rho_0 \rangle_X 
\end{equation}
where $S^{k,j}$ is the $(k,j)$ component of the Gram matrix inverse defined as $S_{k,j} = W_j \times \langle \zeta_k , \zeta_j \rangle_X$. Using eq.~(\ref{resolvingKernel_2}), eq.~(\ref{resolvingKernel_1}) becomes,
\begin{equation}
\label{resolvingKernel_3}
\hat{\rho}(\mathbf{r}) - \rho_0 (\mathbf{r}) = \sum_{k=1}^{K} \zeta_k(\mathbf{r}) \sum_{j=1}^K W_j \times S^{k,j} \times \langle \zeta_j , \rho-\rho_0 \rangle_X.
\end{equation}
The presence of $\langle \zeta_j , \rho \rangle_X$ in the right hand part of this equation indicates that the density distribution actually recovered, $\hat{\rho}(\mathbf{r})$, is assembled from projections of the true unknown density, $\rho(\mathbf{r})$, onto the acquisition kernels. The recovered density is a filtered version of the true density distribution, and the filter (i.e. the resolving kernel) depends on the data. This can be made more explicit by rewriting eq.~(\ref{resolvingKernel_3}) as (Bertero et al., 1985),
\begin{equation}
\hat{\rho}(\mathbf{r}) - \rho_0 (\mathbf{r}) = \int_V \Delta(\mathbf{r},\mathbf{r'}) \times \left( \rho(\mathbf{r'}) - \rho_0 (\mathbf{r'}) \right) \mathrm{d}\mathbf{\mathbf{r'}}, \label{resolvingKernel_4}
\end{equation}
where we introduce the resolving kernel,
\begin{equation}
\Delta(\mathbf{r},\mathbf{r'}) = \sum_{j=1}^{K} b_j (\mathbf{r}) \times \tilde{\zeta}_j(\mathbf{r'}),
\label{resolvingKernel_5a}
\end{equation}
with
\begin{eqnarray}
b_j (\mathbf{r}) & = \sum_{i=1}^K W_i \times \zeta_i(\mathbf{r}) \times S^{i,j} \label{resolvingKernel_5b}, \\
\tilde{\zeta}_j(\mathbf{r'}) & = \int_V w(\mathbf{r'},\mathbf{r''}) \times \zeta_j(\mathbf{r''}) \mathrm{d}\mathbf{\mathbf{r''}}. \label{resolvingKernel_5c}
\end{eqnarray}
$\tilde{\zeta}_j$ is the acquisition kernel $\zeta_j$ modulated by the prior information represented by the $w$ function. For instance, $w$ may be an indicator function used to limit the support of the acquisition kernels to the volume of interest.

\section{Characterisation of the resolving kernel}

A resolving kernel, $\Delta(\mathbf{r},\mathbf{r'})$, is a function defined in the whole space that plays the role of a spatial filter. When applied to the true density distribution, it gives the reconstructed density. The amplitude and the shape of $\Delta$ render the achievable resolution of the reconstructed density structure. According to eq.~(\ref{resolvingKernel_5a}) it is a linear combination of the acquisition kernels. $\Delta$ may be characterized in different ways by using several  properties to quantify its resolution and anisotropy. These properties should be the least possible dependant to a specific resolving kernel and allow the user to easily appreciate the resolution and its eventual bias. In the present study, we simply compare the resolving kernels against the ideal kernel represented by a Dirac distribution $\delta(\mathbf{r}-\mathbf{r'})$. This is achieved through the projection $\gamma$ of $\Delta(\mathbf{r},\mathbf{r'})$ onto a $\delta(\mathbf{r}-\mathbf{r'})$,
\begin{equation}
\gamma(\mathbf{r}) = {\left\langle \Delta(\mathbf{r},\mathbf{r'}) , \delta(\mathbf{r}-\mathbf{r'}) \right\rangle}_X = \int_V w(\mathbf{r},\mathbf{r'}) \times \Delta(\mathbf{r},\mathbf{r'}) \mathrm{d}\mathbf{\mathbf{r'}}
\label{paramGamma}
\end{equation}
where the $X$ scalar product is defined by eq.~(\ref{X_scalar_Product}). Here, the Hamming function is used as the weight function
\begin{equation}
w(\mathbf{r},\mathbf{r'}) = \frac{H_L(\| \mathbf{r}-\mathbf{r'} \|)}{K_w} \times \left[1+\cos\left(\frac{2\pi \times \| \mathbf{r}-\mathbf{r'} \| }{L}\right) \right]
\label{xWeightsHamming}
\end{equation}
where $K_w$ is a normalising constant, $H_L$ is a rectangular pulse that restricts $w$ to $\| \mathbf{r}-\mathbf{r'} \| \in \left[-L/2 ; L/2 \right]$, and $L = 25 \, \mathrm{m}$. If $\mathbf{r}$ or $\mathbf{r'}$ is located outside the volcano we take $w(\mathbf{r},\mathbf{r'}) = 0$. This choice is explained in section \ref{priorInfo}.

We now display resolving kernels corresponding to the data acquisition shown in Fig.~\ref{soufriereCoverage_a} for muon tomography and in Fig.~\ref{soufriereCoverage_b} for gravimetry. Muon radiographies are taken from three sites equidistantly located along the Southern edge of the volcano. Gravity measurements are assumed to be done on a regular grid over the entire lava dome.

\begin{figure}[h]
\centering
\subfigure[vertical observation planes]{
	\includegraphics[width=0.49\textwidth]{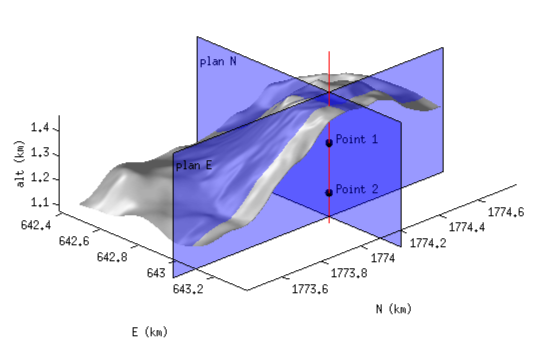}
	\label{situationFigure_a}
}
\hspace{-5mm}
\subfigure[horizontal observation planes]{
	\includegraphics[width=0.49\textwidth]{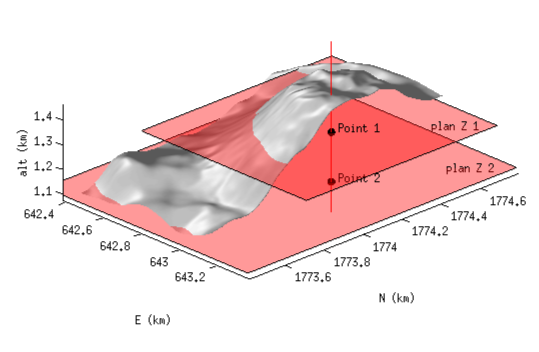}
	\label{situationFigure_b}
}
\caption{3D views of the cross-sections used to represent the resolving kernels in figures \ref{resoKernels1} and \ref{resoKernels2}. The resolving kernels are computed at points $1$ and $2$. Point $1$ is located at a level $Z1$ in the part of the lava dome scanned by the lines of sight of the muon telescope (Fig.~\ref{soufriereCoverage_a}) and point $2$ is located at $Z2$ below the ray coverage of the telescope.}
\label{situationFigure}
\end{figure}

Accounting for the fact that the acquisition kernels $\zeta_k$ are either for gravity or for muon tomography (eq.~(\ref{acquisitionKernelFamily})), we successively consider the case of resolving kernels obtained for muon tomography alone, for gravity data alone, and for a combination of muon tomography and gravity data. We compute $\Delta(\mathbf{r},\mathbf{r'})$ for two positions $\mathbf{r} = \left\lbrace \mathbf{r}_1 ; \mathbf{r}_2 \right\rbrace$ located along a vertical line that goes through the center of the dome (Fig.~\ref{situationFigure}). Points $\mathbf{r}_1$ and $\mathbf{r}_2$ are respectively inside and below the volume of the lava dome spanned by the lines of sight of the telescopes (Fig.~\ref{soufriereCoverage_a}). The parameter $\gamma$ is computed and plotted on the four characteristic slices represented on Fig.~\ref{situationFigure}.

\subsection{Gravimetry kernels}
\label{section_gravimetry_kernels}

Fig.~\ref{acquisitionKernels_b} shows a gravimetry acquisition kernel $\mathcal{G}$. Remind that the data used in the present study are normalized relatively to a reference model with density $\rho_0$ (eq.~(\ref{normalizedGravity})). The gravimetry acquisition kernels are very sensitive to density fluctuations close to the measurement point because of its $1/r^2$ term. The gravity data are actually the component of the gravity field anomaly taken along the local vertical, and the acquisition kernel becomes less and less sensitive as we get closer to the horizontal plane that contains the measurement point.

The gravimetry inverse problem is systematically ill-posed (e.g. Al-Chalabi, 1971) because no matter the number of measurements the resolving kernel mostly integrates information around the measurement positions, i.e. near the surface. An illustration of this problem is given for the resolving kernels of $\mathbf{r}_1$ and $\mathbf{r}_2$ (Fig.~\ref{resoKernels1_c},\ref{resoKernels2_a}). For gravimetry inversions it is more realistic to model $\mathbf{\rho(\mathbf{r})}$ by a function that depends on a few discrete parameters (even if it means losing the linearity between the data and the measurements) rather than trying a continuous inversion.

Observe that $\mathcal{G}$ integrates the density over the entire volume and provide information for point $\mathbf{r}_2$ located below the lines of sight of the telescope.

\subsection{Muon tomography kernels}
\label{section_muon_tomography_kernels}

Fig.~\ref{acquisitionKernels_b} shows a typical muon tomography acquisition kernel $\mathcal{M}$ (eq.~(\ref{acquisitionKernel_2a})). It has a conical shape whose aperture angle depends on the distance between the front and the rear detection matrices of the telescope. The apex of the kernel is located at the telescope, and the kernel widens as we move away from the telescope thus the local sensitivity is decreasing. Moreover the triangular shape of the intra-pixel acceptance $\mathcal{P}_m$ (see Fig.~\ref{Figure_Frame} and Fig.~\ref{Figure_Pixel_Acceptance}) makes the sensitivity maximum along the main line of sight $\mathbf{s}_m$.

The Fig.~\ref{acquisitionKernels_b} shows we are as sensitive to a density change occurring on a few tenth of meters in front of the telescope as to the same change happening on a few hundred of meters beside the volcano. It reveals how deterministic is the telescope position. If one desires to image or monitor a specific region belonging to a bigger structure, the measurement will be much more sensitive if the telescope is in front of it. The important heterogeneities inside the muon tomography acquisition kernels forbid us to use the Radon transform mathematical corpus. For an equivalent resolution and scanning the kernels can be regularized taking the telescope away from the volcano and reducing the angular aperture. We then get into the typical experimental conditions of a medical X-ray tomography. But the consequences are a weaker particle flux (a longer acquisition time) and a greater sensitivity to potential noises. So a compromise has to be found, but the actual lack of understanding of the noises and the already very long acquisition times we are facing lead us to take the telescope the closest we can to the volcano.

We draw the reader's attention to the fact that, despite their compact support, the acquisition kernels overlap each others for neighbour main lines of sight (see Fig.~\ref{Figure_Pixel_Acceptance}). As will be seen below, this characteristic is fundamental to understand the shape of the resolving kernels.

A muon tomography resolving kernel is a linear combination of muon tomography acquisition kernels $\mathcal{M}$ (eq.~(\ref{resolvingKernel_5a})), and the inversion process optimizes the $b_i$ coefficients to obtain the best density model (eq.~(\ref{resolvingKernel_5b})). Fig.~\ref{resoKernels1_a} and Fig.~\ref{resoKernels1_b} show the resolving kernel for point $\mathbf{r_1}$ for different combinations of tomography datasets.

When using data acquired from a single place located at the Southernmost edge of the volcano (Ravine Sud, see Fig.~\ref{situationFigure_a}), the resolving kernel (Fig.~\ref{resoKernels1_a}) encompasses lines of sight spanning a limited range of azimuths. Consequently, the filtering effect of $\Delta(\mathbf{r}_1,\mathbf{r'})$ integrates $\rho$ along a long narrow cone to give the estimated density $\hat{\rho}(\mathbf{r}_1)$. The fact that $\Delta(\mathbf{r}_1,\mathbf{r'})$ has not a compact support like the $\mathcal{M}$ kernels comes from the overlapping of neighbour acquisition kernels that produces a transfer of information among lines of sight.

When simultaneously using all three muon tomography sites, the resolving kernel includes acquisition kernels that span a wider range of sight azimuths. Consequently, the resolving kernel is more localized onto point $\mathbf{r}_1$ (Fig.~\ref{resoKernels1_b}). However, the number of radiographies remains small and the kernel has a spider shape visible in the horizontal slice at the far right of Fig.~\ref{resoKernels1_b}.

Observe that the resolving kernel $\Delta(\mathbf{r}_2,\mathbf{r}) = 0$ since all acquisition kernels are null in this part of the volcano.

\subsection{Joined muon tomography and gravimetry kernels}

We now consider resolving kernels computed by using both muon $\mathcal{M}$ and gravity $\mathcal{G}$ acquisition kernels. 

\begin{figure}[h]
\centering
\subfigure[gravimetry data]{
	\includegraphics[width=0.48\textwidth]{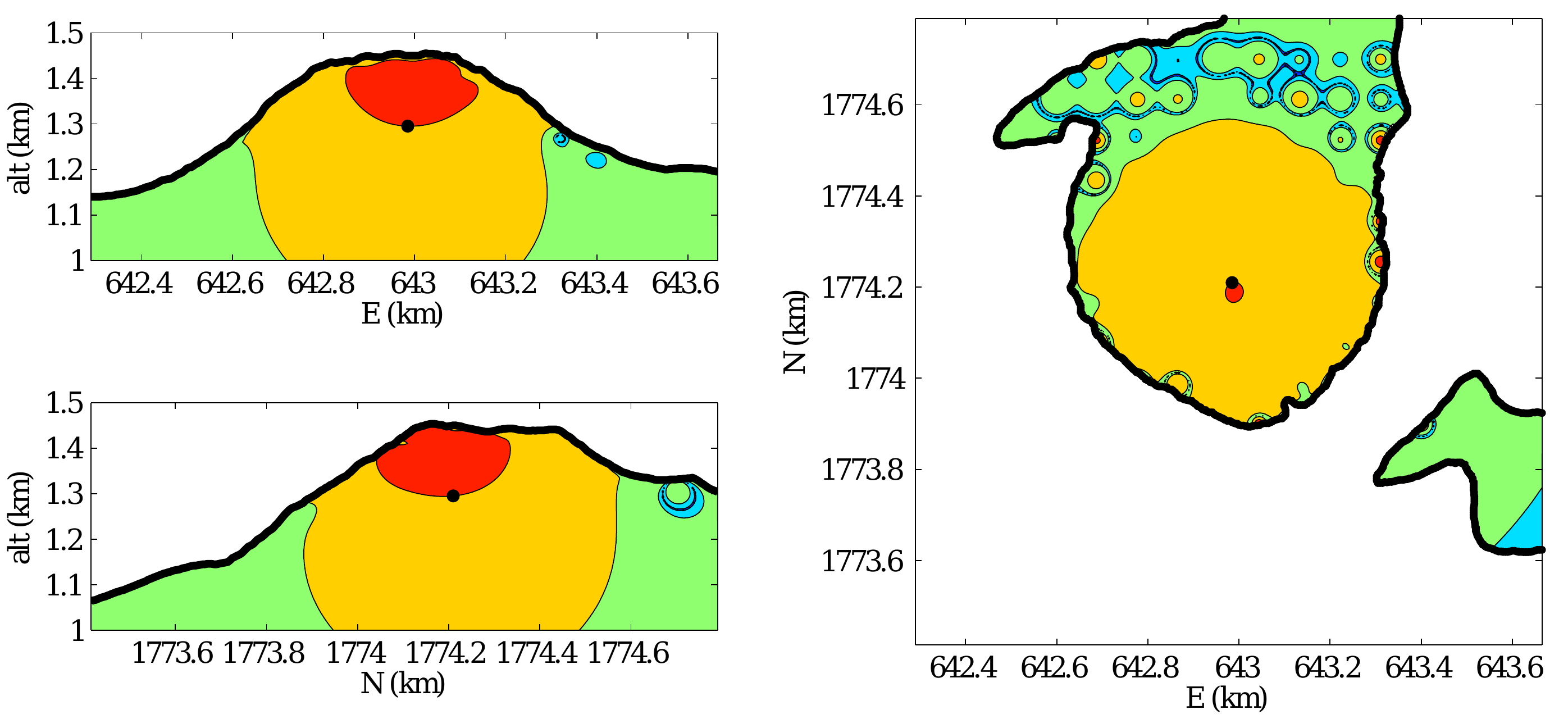}
	\label{resoKernels1_c}
}
\subfigure[Ravine Sud tomography data]{
	\includegraphics[width=0.48\textwidth]{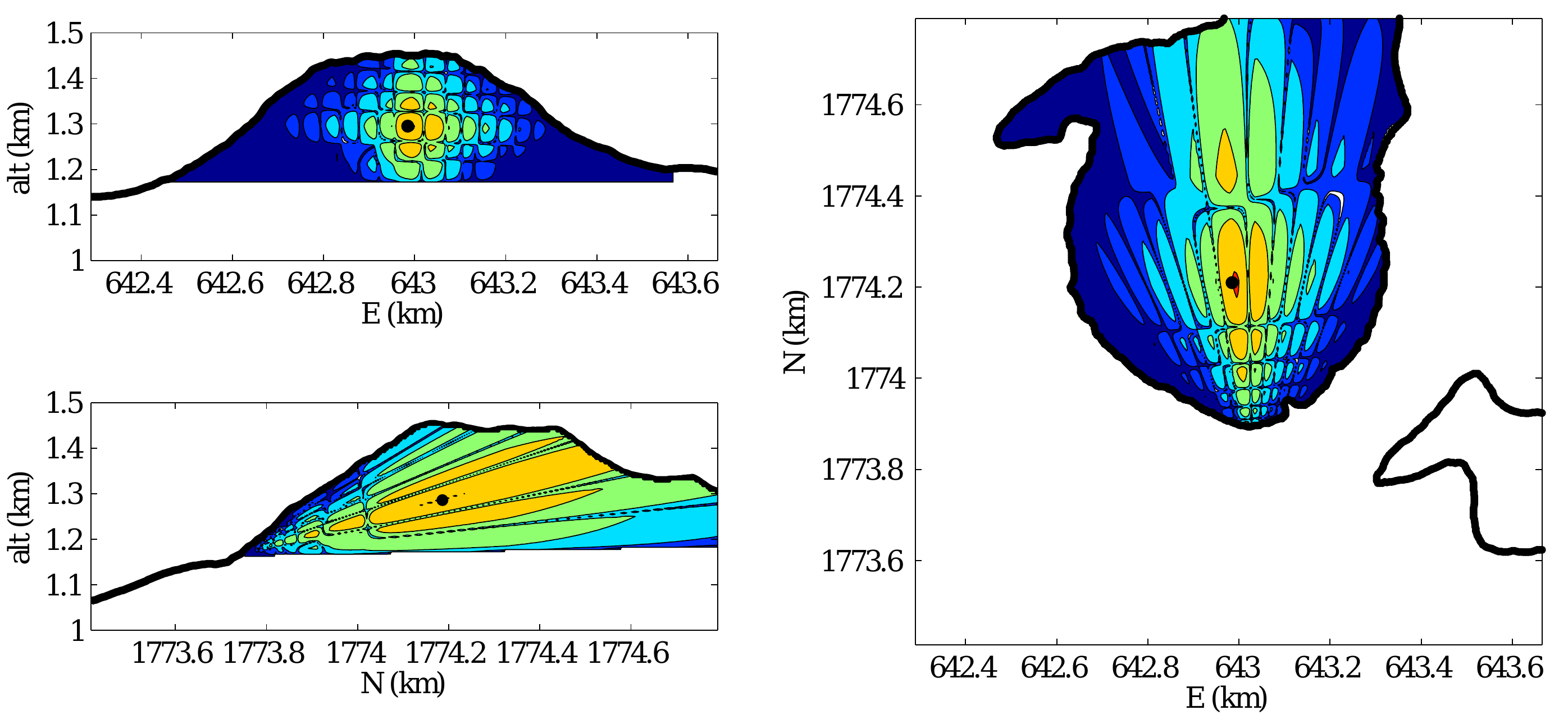}
	\label{resoKernels1_a}
}
\subfigure[all sites tomography data]{
	\includegraphics[width=0.48\textwidth]{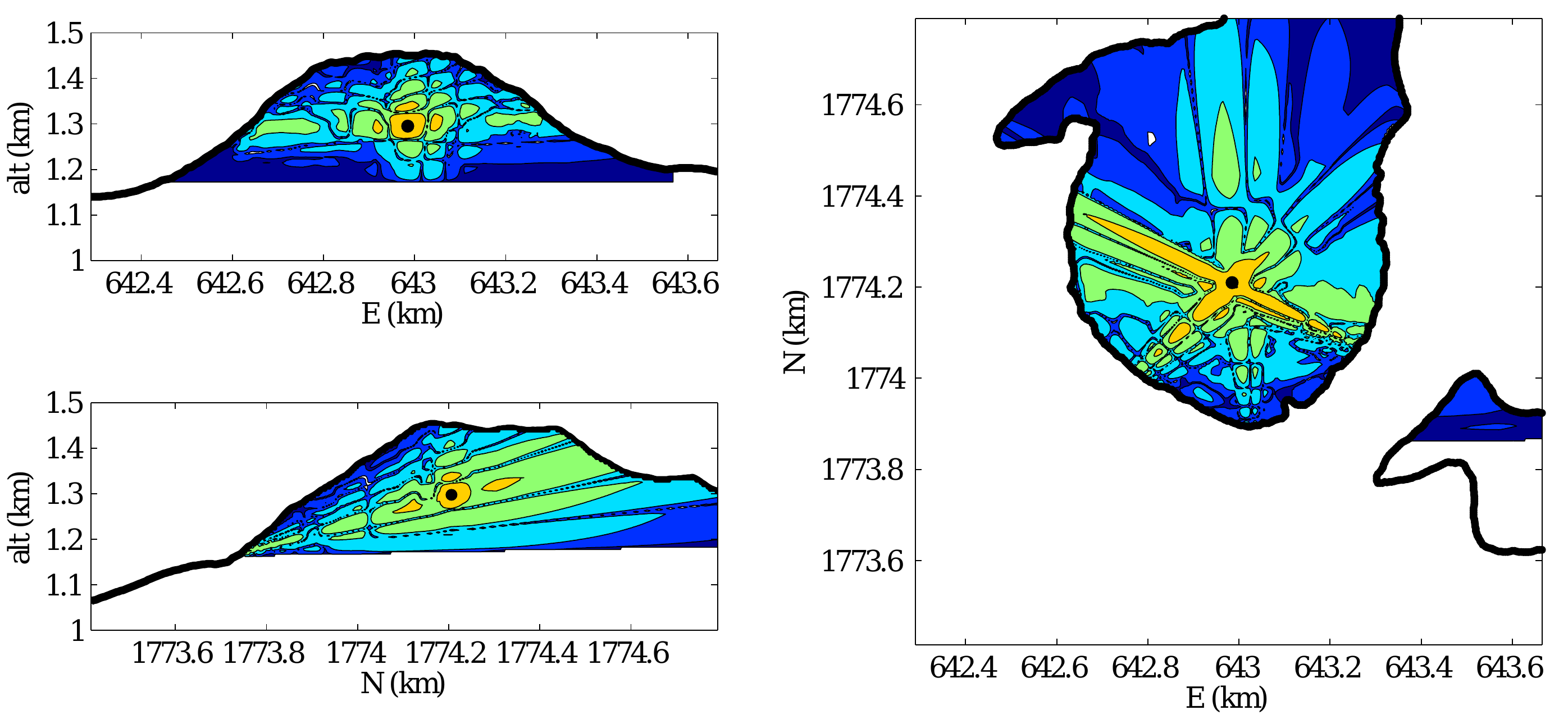}
	\label{resoKernels1_b}
}
\subfigure[gravimetry and tomography data]{
	\includegraphics[width=0.48\textwidth]{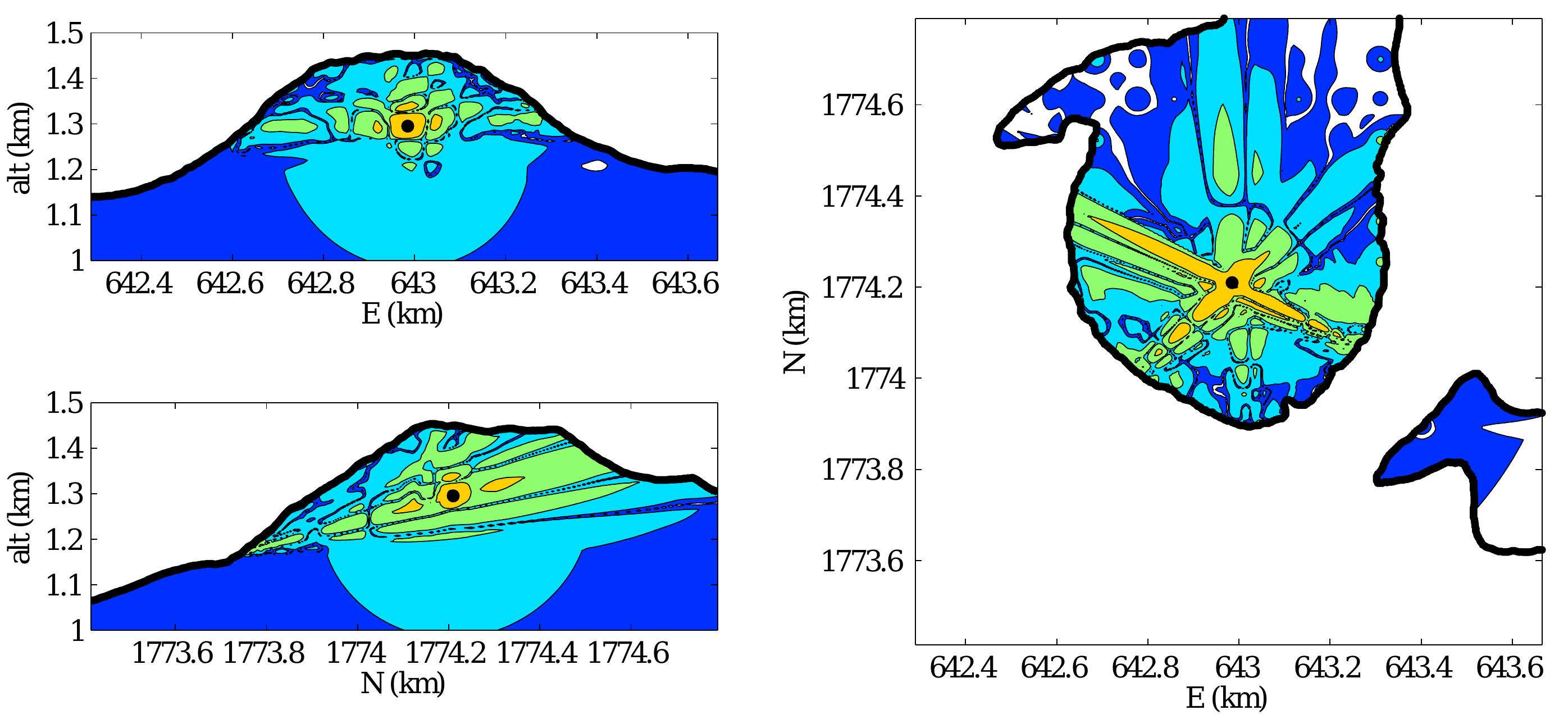}
	\label{resoKernels1_d}
}
\subfigure[]{
	\includegraphics[width=0.9\textwidth]{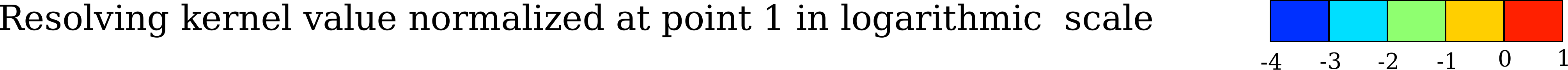}
}
\caption{Resolving kernel at point $\mathbf{r_1}$ (Fig.~\ref{situationFigure}) obtained for the gravity data alone, \textit{(a)}, the Ravine Sud muon tomography alone, \textit{(b)}, all three muon radiographies , \textit{(c)}, and the joined muon and gravity datasets \textit{(d)}. See Fig.~\ref{soufriereCoverage_b} for the locations of gravity measurements and the three sites for muon radiographies. The resolving kernel absolute value is normalized with reference to the value computed at point $\mathbf{r_1}$ and represented with a $\log_{10}$ scale.}
\label{resoKernels1}
\end{figure}

Gravimetry does not improve significantly the inversion process at $\mathbf{r}_1$, and the resolving kernel $\Delta(\mathbf{r}_1,\mathbf{r'})$ (Fig.~\ref{resoKernels1_d}) looks very similar to the one obtained for the muon radiographies alone (top right of Fig.~\ref{resoKernels1_b}). The information provided by muon tomography is dominant relative to  gravimetry excepted at the immediate vicinity of the gravity measurement points.

\begin{figure}[h]
\centering
\subfigure[gravimetry data]{
	\includegraphics[width=0.48\textwidth]{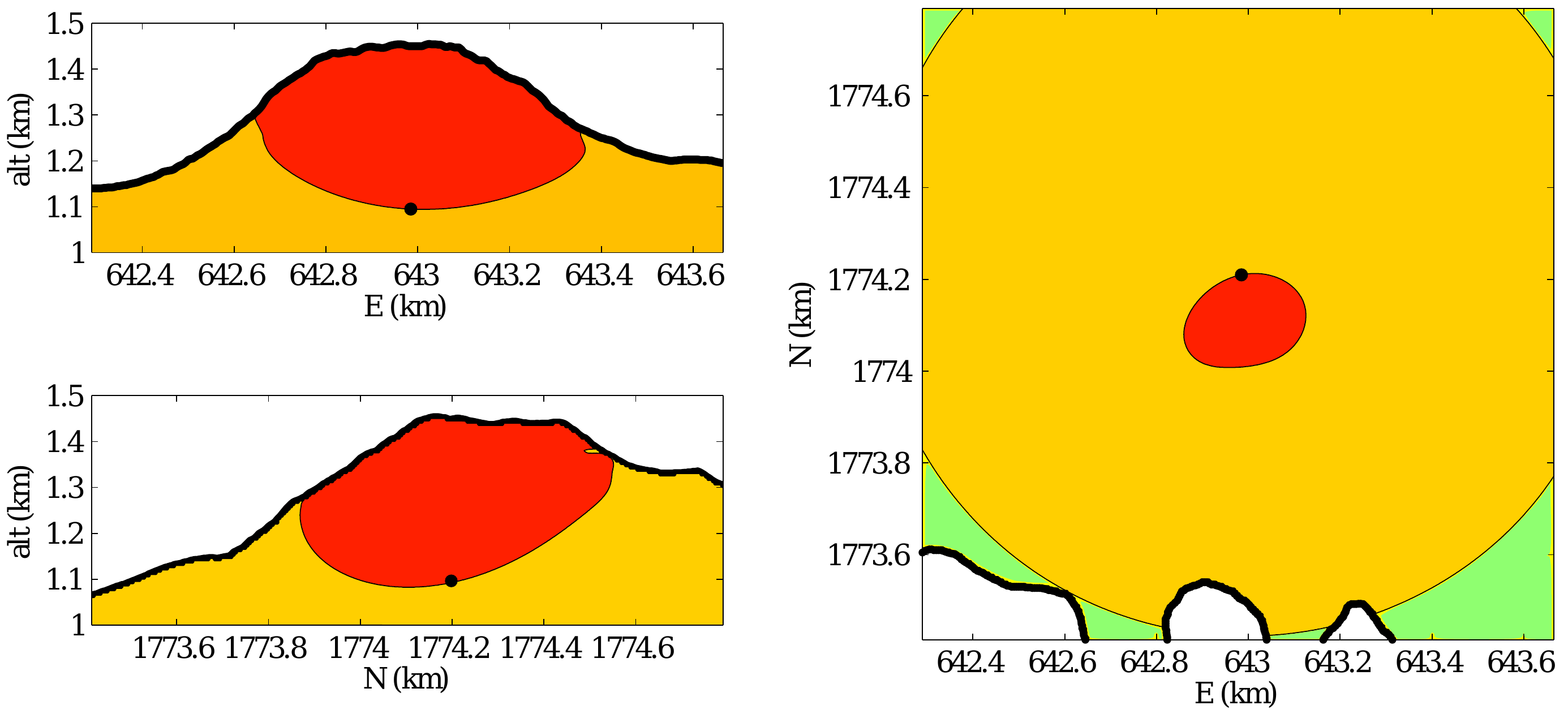}
	\label{resoKernels2_a}
}
\subfigure[gravimetry and tomography data]{
	\includegraphics[width=0.48\textwidth]{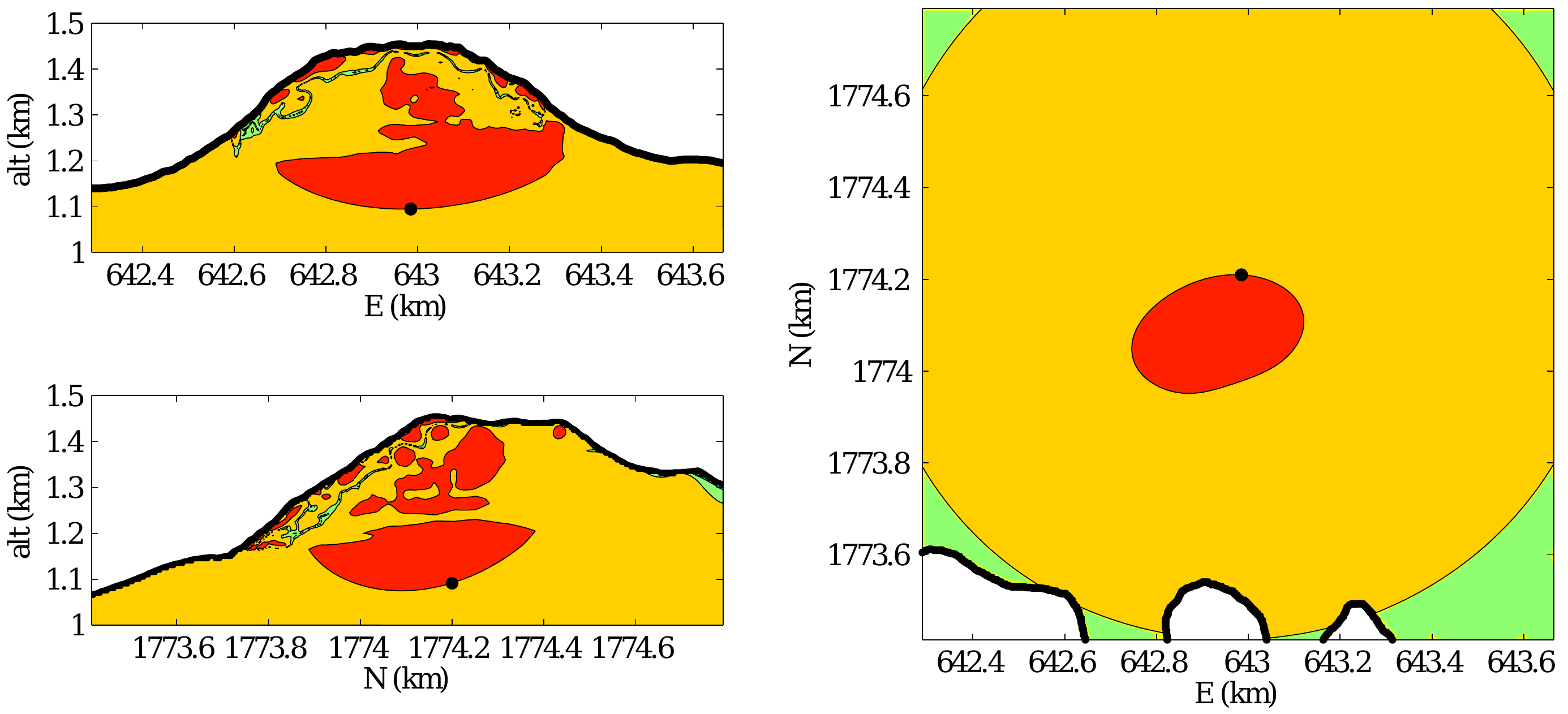}
	\label{resoKernels2_b}
}
\subfigure[gravimetry and tomography data + prior information]{
	\includegraphics[width=0.48\textwidth]{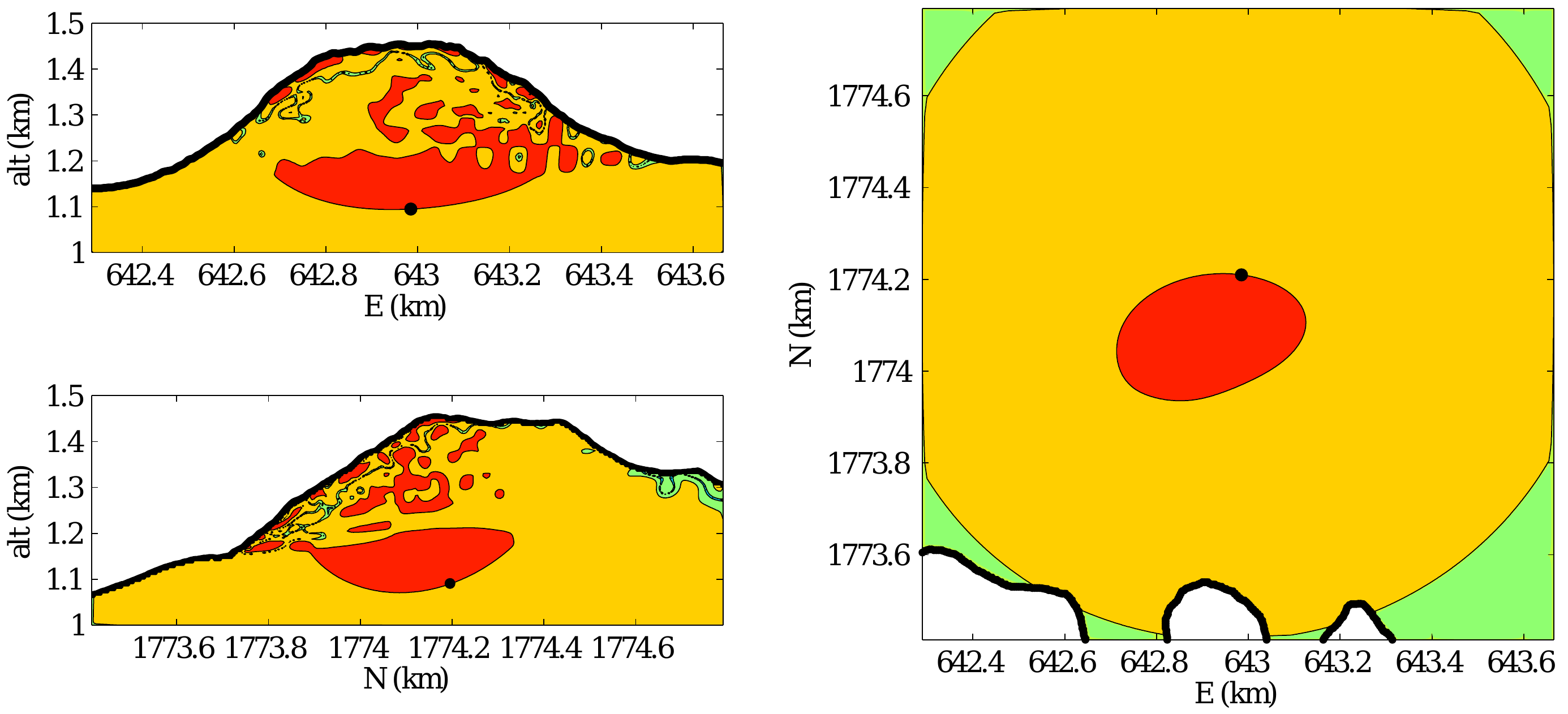}
	\label{resoKernels2_c}
}
\subfigure[]{
	\includegraphics[width=0.9\textwidth]{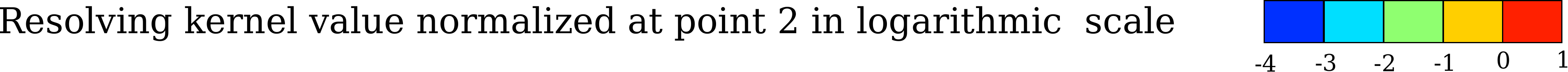}
}
\caption{Resolving kernel at point $\mathbf{r_2}$ (Fig.~\ref{situationFigure}) obtained for the gravity data alone \textit{(a)} and for the joined muon and gravity datasets \textit{(b)}. Figure \textit{(c)} is obtained with both muon and gravity datasets and some prior information about the density spatial correlation. See Fig.~\ref{soufriereCoverage_b} for the locations of gravity measurements and the three sites for muon radiographies. The resolving kernel absolute value is normalized with reference to the value computed at point $\mathbf{r_2}$ and represented with a $\log_{10}$ scale.}
\label{resoKernels2}
\end{figure}

The situation is very different for point $\mathbf{r}_2$ where the resolving kernel (Fig.~\ref{resoKernels2_b}) obtained by joining muon radiographies and gravity data appear very different from the gravity kernel (Fig.~\ref{resoKernels2_a}). The most conspicuous effect is that muon data compensates the great sensitivity of gravimetry at near-surface locations by shifting the center of mass of the resolving kernel downward. This considerably improves the vertical resolution achievable in the deepest parts of the volcano.

The conclusions are different if only one tomography acquisition is available. In that case the gravimetry measurements have an impact on the upper part of the dome because they contribute to resolve the ambiguity about the anomaly spatial depth relatively to the acquisition position. But then the zone below the dome will lack data to be properly constrained.

\subsection{Impact of prior information}
\label{priorInfo}

The choice made for the $X$ and $Y$ weight functions $w(\mathbf{r'},\mathbf{r''})$ (eq.~\ref{X_scalar_Product}) and $W_{i = 1 ... K}$ (eq.~(\ref{resolvingKernel_1b})) has an important influence on the obtained resolving kernels $\Delta(\mathbf{r},\mathbf{r'})$.

In the $X$ space, the diagonal term $w(\mathbf{r'},\mathbf{r''}=\mathbf{r'})$ permits to adjust the local degree of prior knowledge on $\rho(\mathbf{r})$. For instance, $w(\mathbf{r},\mathbf{r}) = 0$ in regions where $\rho(\mathbf{r})$ is assumed sufficiently well-known to have no impact on our measurements. This corresponds to situations where $\rho_0(\mathbf{r}) = \rho(\mathbf{r})$ and where the concerned regions have not to be accounted for in the inversion process. In our case we use it to cancel the free-air impact on muon tomography and gravimetry, but we can also constrain it to incorporate direct field measurements of the density.

The non-diagonal part $w(\mathbf{r'},\mathbf{r''} \ne \mathbf{r'})$  may be used to introduce a spatial correlation in $\rho$. This can be done through $\tilde{\zeta}$ which the convolution of $\zeta$ with $w$ (eq.~(\ref{resolvingKernel_5c})). Here, we use a simple Hamming function with a $25~\textrm{m}$ correlation length everywhere in the dome (eq.~(\ref{xWeightsHamming})), and $\tilde{\zeta}$ is a smoothed version of $\zeta$ which attenuates the $1/r^2$ effect previously mentioned (for muon tomography it permits to get closer to the X-ray tomography experimental conditions previously detailed). The correlation introduced by the Hamming function increases the acquisition kernels sensitivity further from the measurement point toward the central and the Northern parts of the dome. This produces a better localization of the resolving kernel at $\mathbf{r_2}$ as can be checked by comparing Fig.~\ref{resoKernels2_c} with Fig.~\ref{resoKernels2_b} where no spatial correlation was applied. The counterpart of this effect is a de-sharpening of the kernel at point $\mathbf{r_2}$. $w$ is a regularizing low-pass filter that removes spurious short-wavelength fluctuations in the density model and reduces the ill-conditioning of the inverse problem (e.g. Bertero et al., 1988).

The choice of $w$ is problem-dependent and must be sustained by prior knowledge. The Hamming function acts as a low pass filter with a limited support compatible with the large homogeneous zones  observed on the field: massive andesite, hydrothermally altered material and possibly large cavities.

In the $Y$ space, the weights $W_{i = 1 ... K}$ allow to assign different quality factors to the available data at one inversion location. For instance, in muon tomography, the $W'\mathrm{s}$ permit to account for the fact that all observation axes have not the same integrated acceptance $\mathcal{T}_m$ (Fig.~\ref{Figure_Pixel_Acceptance}). The quality of the gravity data strongly depends on the ground stability (i.e. tilt stability during measurement sequences) and the presence of wind (i.e. vibrations of the gravity-meter). The non-diagonal terms $W_{ij ~,~ i \neq j}$ are null as the measurements are independent the ones from the others.

\subsection{$\gamma$ maps}

\begin{figure}[h]
\centering
\subfigure[gravimetry data]{
	\includegraphics[width=0.48\textwidth]{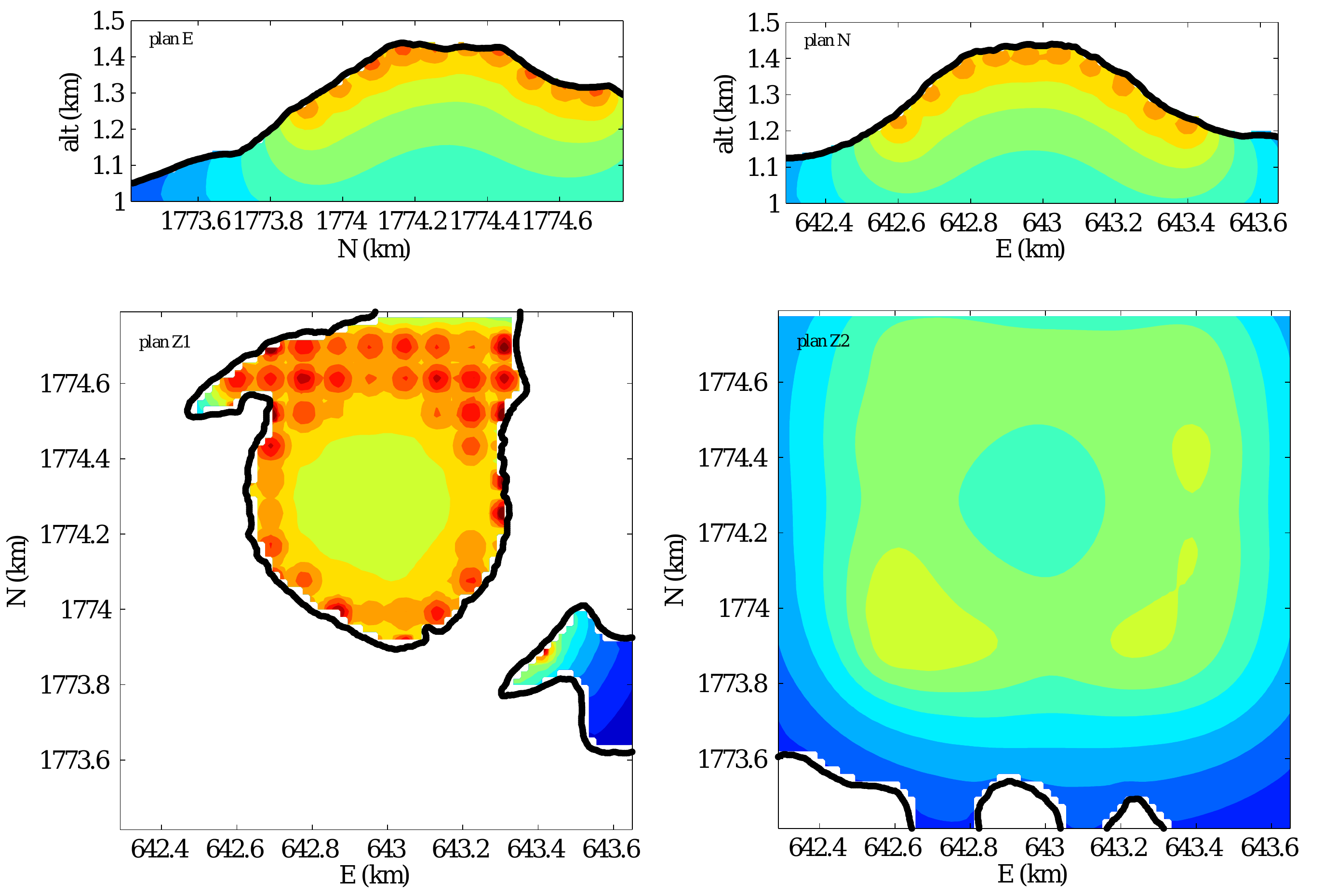}
	\label{gammaMap_a}
}
\subfigure[gravimetry and tomography data]{
	\includegraphics[width=0.48\textwidth]{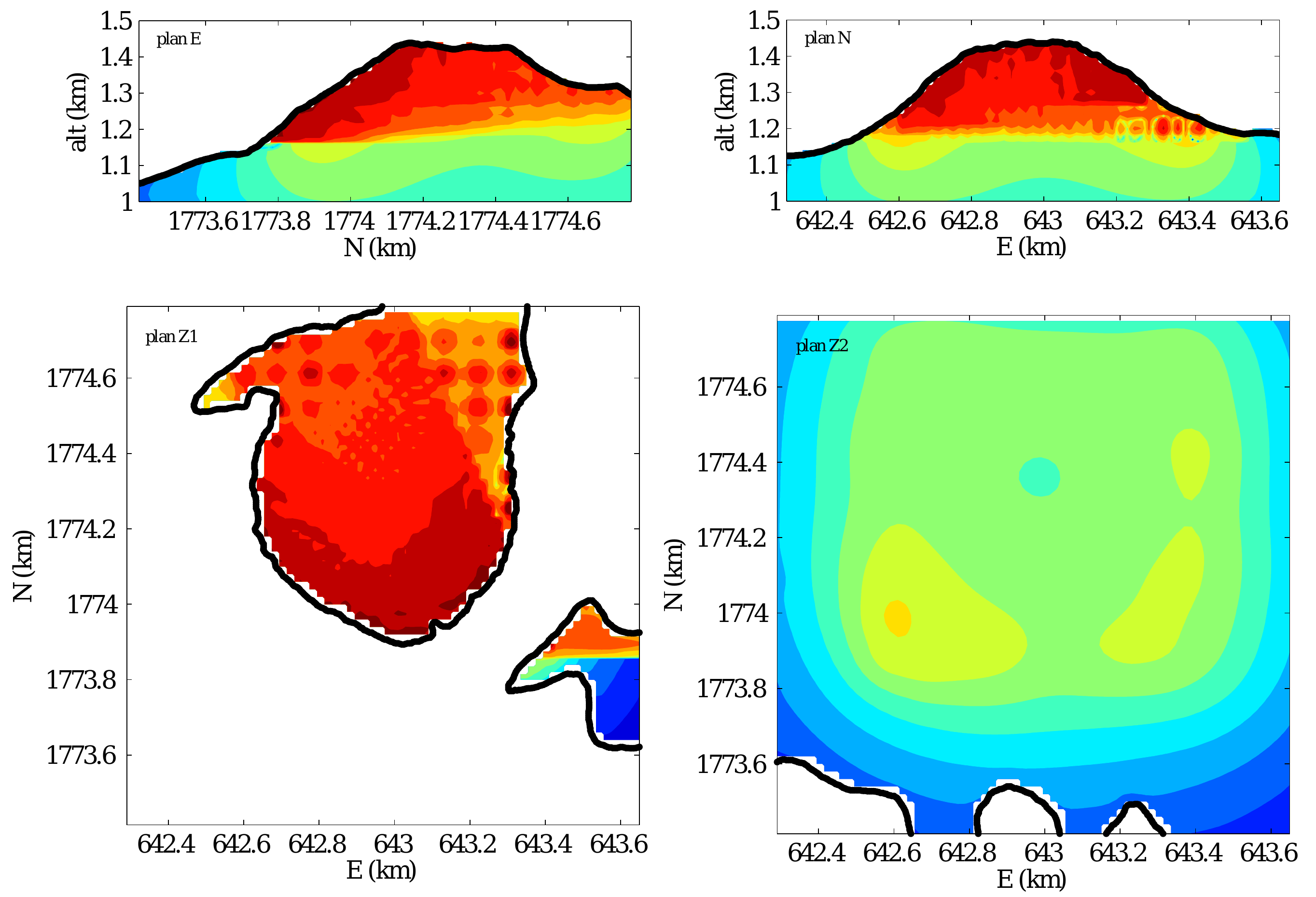}
	\label{gammaMap_b}
}
\subfigure[]{
	\includegraphics[width=0.9\textwidth]{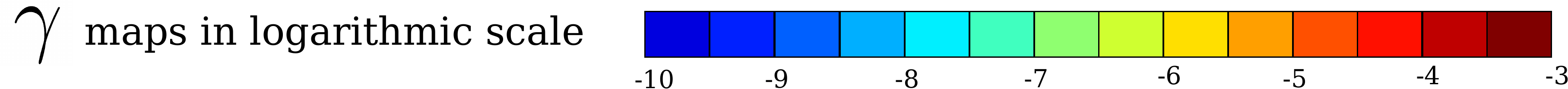}
}
\caption{Representation of $| \gamma |$ on the four slices defined in Fig. \ref{situationFigure}. The results are represented with a $\log_{10}$ scale.}
\label{gammaMap}
\end{figure}

The $\gamma(\mathbf{r})$ index defined in eq.~(\ref{paramGamma}) may be used to estimate the resolution achievable everywhere in the volcano. Fig.~\ref{gammaMap} shows slices of the $\gamma$ function obtained for the gravity data (left part of Fig.~\ref{gammaMap}) and by joining the three muon tomography data sets together with the gravity measurements (Fig.~\ref{gammaMap_b}). The gravimetry $\gamma$ slices clearly reveal the important sensitivity of the data to density variations located in the immediate vicinity of the measurement points and the very low sensitivity to density structure located deeper in the lava dome.

The muon-gravimetry $\gamma$ slices confirm the results obtained for points $\left\lbrace \mathbf{r}_1 ; \mathbf{r}_2 \right\rbrace$ and show the considerable improvement of the resolution obtained when jointly using the muon and gravity datasets. They also reproduce the asymmetric resolution due to the conical shape of the muon acquisition kernels $\mathcal{M}$. Since the places occupied by the telescope are located along the Southern edge of the volcano, a finer resolution is obtained for the Southern part of the lava dome. This corresponds to the dark-red circular sector visible in the upper horizontal slice in Fig.~\ref{gammaMap_b}. As expected, the resolution is coarser in the central and Northern parts of the dome. The same slice shows that the North-Eastern region of the volcano is resolved by the gravity data alone since no lines of sight of the telescope cross this part of the dome.

The $\gamma$ map is a useful tool to plan an acquisition survey. One can easily compute how $\gamma$ is changed with different possible measurement campaigns and select the most pertinent one depending on the region of interest, the available time on the field and the accessibility of the site. This choice is critical as muon tomography acquisitions are long (a few weeks) and gravity measurements delicate. It can also be used to design a mesh for the problem. The meshing elements density can roughly follows the $\gamma$  map fluctuations.

We emphasize that other definitions may be used for $\gamma$ and that a single index may prove insufficient to characterize the shape of resolving kernels. Consequently, we recommend to perform a 3D examination of individual resolving kernels at locations of particular importance (i.e. like detecting places where density changes occur).

\section{Conclusion}

The resolving kernel analysis discussed in the present study allows to quantitatively assess the way by which gravity data and density muon radiographies may be joined to reconstruct the density distribution inside geological bodies. A main result concerns the improvement of the resolution obtained in the deep regions of the density model when joining muon and gravity data, and despite the fact that these regions are not sampled by muon tomography. Part of the information brought by the muon data is transferred to the deep regions of the model through the long-range coupling of the gravity acquisition kernels (Fig.~\ref{acquisitionKernels_a}) used to construct the joined resolving kernels (eq.~(\ref{resolvingKernel_5a})).

The compact support of the muon acquisition kernels (Fig.~\ref{acquisitionKernels_b}) allows muon tomography to obtain high-resolution models of the density structure of La Soufri\`ere but restricted to the upper part of the dome. Gravimetry cannot be used on its own to make a proper continuous inversion because it is too sensitive to the very close subsurface. Therefore coupling the two techniques does not lead to significant improvement of the density model in regions where muon data are available (compare Fig.~\ref{resoKernels1_b} and Fig.~\ref{resoKernels1_d}).

The muon tomography acquisition kernel has a conic shape. and noise considerations obligate us to put the cone apex just in front the studied structure. It results in a large decrease of the sensitivity between the front and the rear of the volcano. This problem adds to the heterogeneous tomography sampling and forbid us to use the standard Radon transform usually adopted in X-ray tomography medical experiments to inverse the density. It also shows how deterministic is the telescope position if one desires to image or monitor a specific region belonging to a bigger body.

The positive weight function $w$ of the inner product (eq.~(\ref{X_scalar_Product})) may be used to introduce prior information both by limiting the support of the density distribution to reconstruct and by introducing a spatial correlation smoothing the $1/r^2$ effect of muon tomography and gravimetry acquisition kernels. This extends the range of sensitivity of the measurements and results in a solution with a more homogeneous quality. 

Finally the $\gamma$ maps give an overview of the resolving kernel geometry everywhere in the dome which can easily be used to optimally plan future acquisition surveys with respect to data already available.

\section*{acknowledgments}
This study is part of the \textsc{anr-14-ce04-0001 diaphane} project. We acknowledge the financial support from the UnivEarthS Labex program of Sorbonne Paris Cit\'e (\textsc{anr-10-labx-0023} and \textsc{anr-11-idex-0005-02}). This study also received funding from Swisstopo through the Mont-Terri MD Project. This is IPGP contribution ****.

\end{document}